\documentclass[review,authoryear,12pt]{elsarticle}

\usepackage{graphicx}
\graphicspath{ {./Figures/} }
\usepackage[tight,footnotesize, nooneline]{subfigure}

\usepackage{adjustbox} 
\usepackage[hidelinks]{hyperref} 

\usepackage{amssymb}
\usepackage{amsthm}
\usepackage{amsmath}

\usepackage{lineno}
 
\usepackage{multirow}

\journal{arXiv}

\newcommand\copyrighttext{%
   \textcopyright\ 2016. This manuscript version is made available under the CC-BY 4.0 license \url{https://creativecommons.org/licenses/by/4.0/}}

\begin{document}

\begin{frontmatter}



\title{Nonlinear 3-D simulation of high-intensity focused ultrasound therapy in the kidney}





\author[Affil1]{Visa Suomi \corref{cor1}}
\author[Affil2]{Jiri Jaros}
\author[Affil3]{Bradley Treeby}
\author[Affil1]{Robin Cleveland}
\address[Affil1]{Department of Engineering Science, University of Oxford, Parks Road, Oxford, OX1 3PJ, UK}
\address[Affil2]{Faculty of Information Technology, Brno University of Technology, Brno, Czech Republic}
\address[Affil3]{Department of Medical Physics and Biomedical Engineering, University College London, Wolfson House, 2-10 Stephenson Way, London, NW1 2HE, UK}
\cortext[cor1]{Corresponding Author: Visa Suomi, Institute of Biomedical Engineering, Old Road Campus Research Building, University of Oxford, Oxford OX3 7DQ, UK; Email, visa.suomi@eng.ox.ac.uk; Phone, +44 (0) 1865 617660}

\begin{abstract}
Kidney cancer is a severe disease which can be treated non-invasively using high-intensity focused ultrasound (HIFU) therapy. However, tissue in front of the transducer and the deep location of kidney can cause significant losses to the efficiency of the treatment. The effect of attenuation, refraction and reflection due to different tissue types on HIFU therapy of the kidney was studied using a nonlinear ultrasound simulation model. The geometry of the tissue was derived from a computed tomography (CT) dataset of a patient which had been segmented for water, bone, soft tissue, fat and kidney. The combined effect of inhomogeneous attenuation and sound-speed was found to result in an 11.0 dB drop in spatial peak-temporal average (SPTA) intensity in the kidney compared to pure water. The simulation without refraction effects showed a 6.3 dB decrease indicating that both attenuation and refraction contribute to the loss in focal intensity. The losses due to reflections at soft tissue interfaces were less than 0.1 dB. Focal point shifting due to refraction effects resulted in $-$1.3, 2.6 and 1.3 mm displacements in x-, y- and z-directions respectively. Furthermore, focal point splitting into several smaller subvolumes was observed. The total volume of the secondary focal points was approximately 46\% of the largest primary focal point. This could potentially lead to undesired heating outside the target location and longer therapy times. 
\end{abstract}


\end{frontmatter}

\copyrighttext

\pagebreak









\section*{Introduction}

Kidney cancer is the 13$^{\textrm{th}}$ most common cancer in the world with approximately 338,000 cases diagnosed in 2012 of which 214,000 were in men and 124,000 in women \citep{globocan2012cancer}. In the same year approximately 143,000 people died due to the disease. Early diagnosis as well as safe and effective therapy methods are therefore crucial for the survival of patients. Typically kidney cancer is treated surgically which is effective \citep{van2011prospective}, but this can lead to complications in as many as 19\% of cases \citep{gill2003comparative}. Alternative, minimally invasive, therapies such as cryotherapy \citep{gill2005renal} and radiofrequency ablation \citep{gervais2005radiofrequency} reduce the risk of complications and often result in shorter hospital stays. However, neither of these methods is completely non-invasive and therefore still contain a risk of infection, seeding metastases and other complications.

High-intensity focused ultrasound (HIFU) is a non-invasive therapy method which does not require puncturing the skin and typically has minimal or no side-effects. HIFU therapy can be used clinically to treat cancerous tissue in kidney, but the oncological outcomes have been variable \citep{ritchie2010extracorporeal}. This has been thought to be partly due to the attenuation properties of peri-nephric fat \citep{ritchie2013attenuation} which results in poor delivery of HIFU energy to the target focal point. The effect of attenuation might be significant especially in the nonlinear case where higher harmonic frequencies generated during HIFU therapy are strongly attenuated. In addition to attenuation, the defocusing of ultrasound due to refraction and the reflections at the tissue interfaces might result in significant loss of HIFU energy \citep{illing2005safety}.

The aim of this research is to investigate how the attenuation, reflection and refraction effects of different tissue types affect the overall efficacy of HIFU therapy of the kidney. This was done by performing nonlinear HIFU therapy simulations in a segmented computed tomography (CT) dataset of a patient in 3-D.

\section*{Simulations}

\subsection*{Parallelised nonlinear ultrasound simulation model}

The HIFU simulations were calculated using the parallel k-Wave toolbox. The k-Wave toolbox models ultrasound wave propagation in soft tissue using a generalised version of the Westervelt equation which accounts for nonlinearity, material heterogeneities and power law absorption. The governing equations are solved using a k-space pseudospectral approach where the Fourier collocation spectral method is used to calculate spatial gradients, and a k-space corrected finite difference scheme is used to integrate forwards in time.

The toolbox is designed for deployment on large distributed computer clusters with thousands of compute cores \citep{nikl2014parallelisation}. The simulation domain is partitioned over one or two dimensions and distributed among the cores. Since the gradient calculation requires the fast Fourier transform (FFT) to be calculated over the whole domain, global data exchange is performed in each simulation time step. Although this has been proven to be a bottleneck, the code efficiency remains acceptable up to to 8192 compute cores \citep{nikl2014parallelisation}. The simulation data sampling and storing is performed via a parallel I/O module based on the HDF5 library and Luster file system.


\subsection*{Simulation geometry and execution}

\begin{figure*}[htbp!]
	\vspace{2mm}
    \centering
    \subfigure[]
    {
        \includegraphics[width=0.3\textwidth]{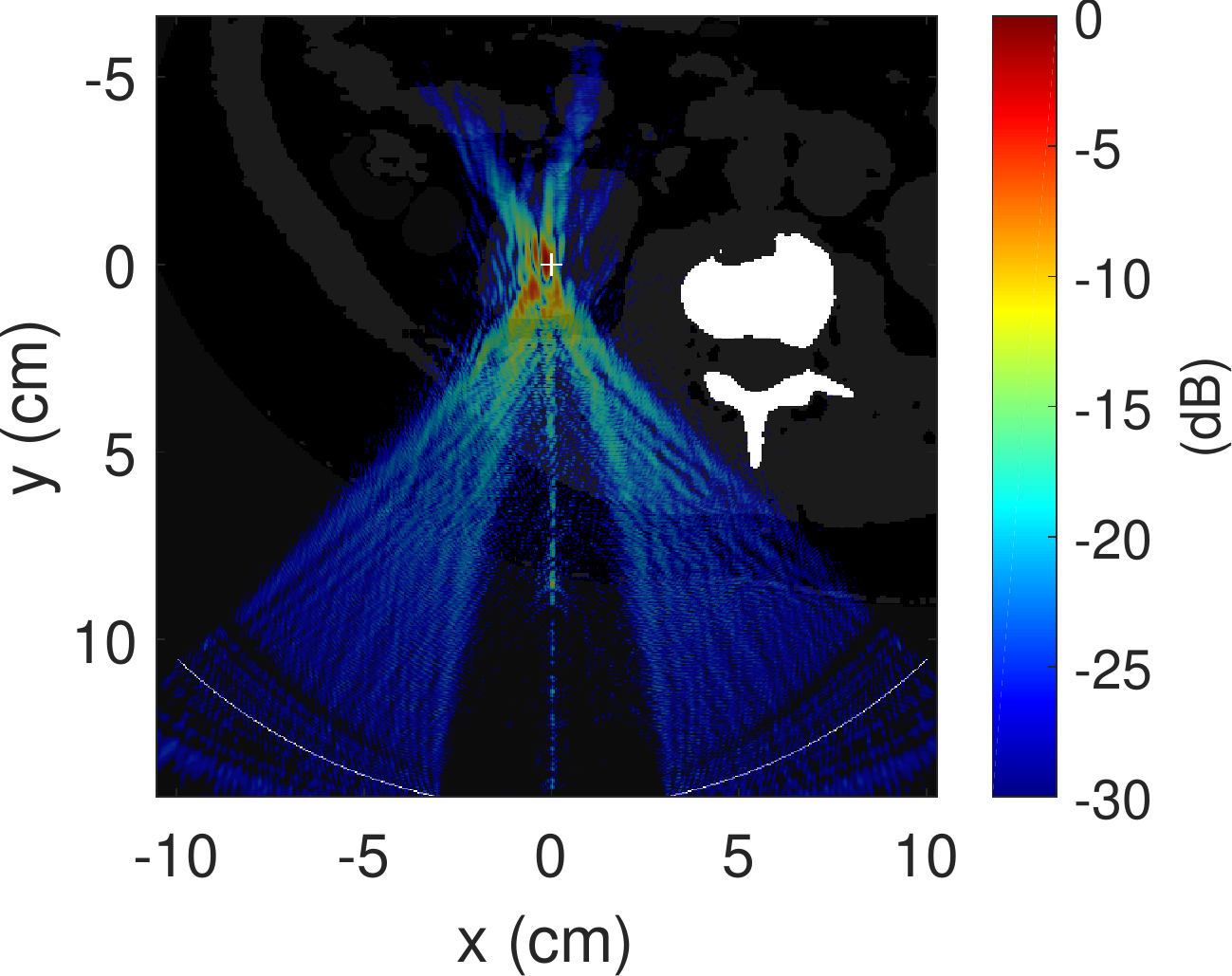}
    }
    \subfigure[]
    {
        \includegraphics[width=0.3\textwidth]{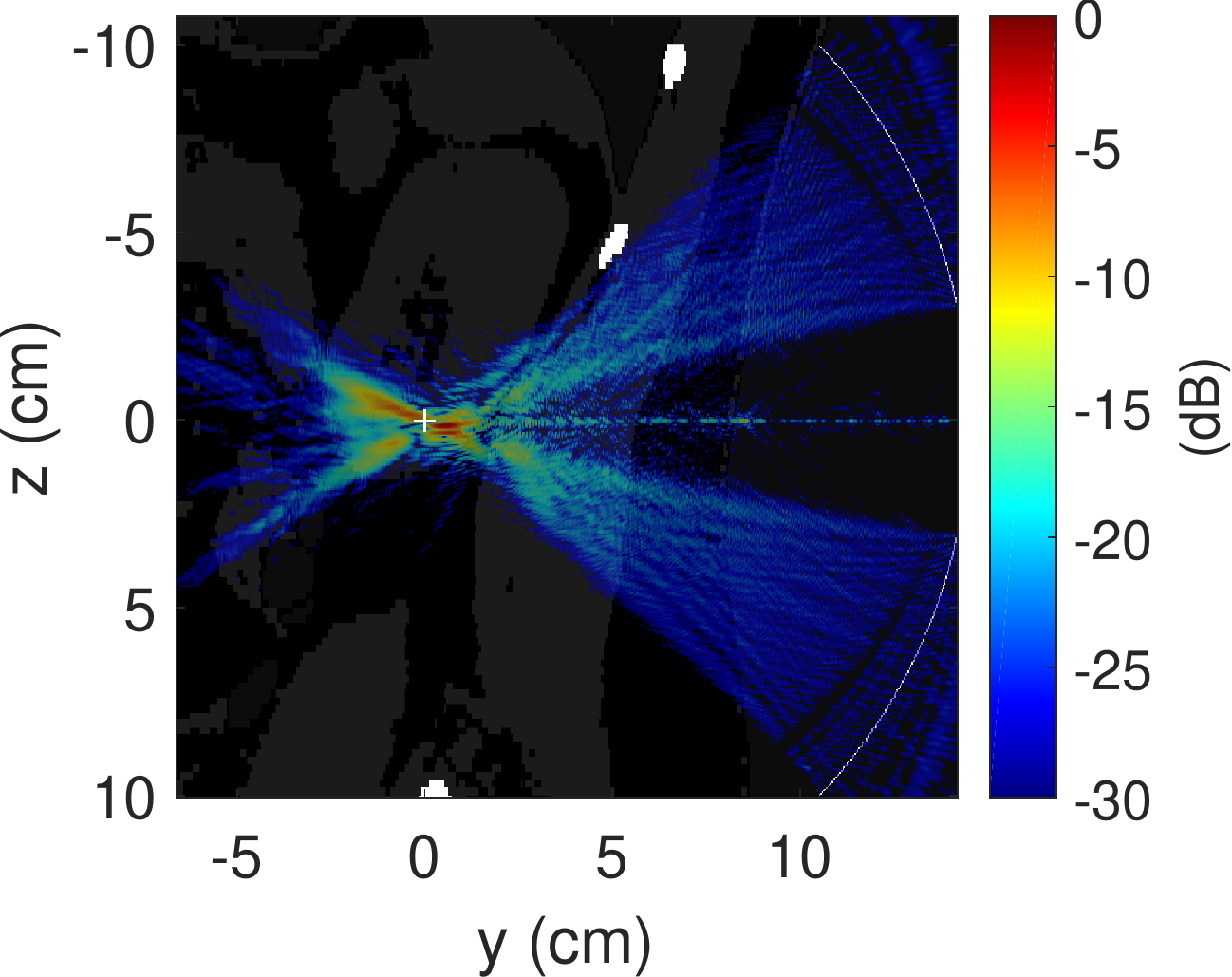}
    }
    \subfigure[]
    {
        \includegraphics[width=0.3\textwidth]{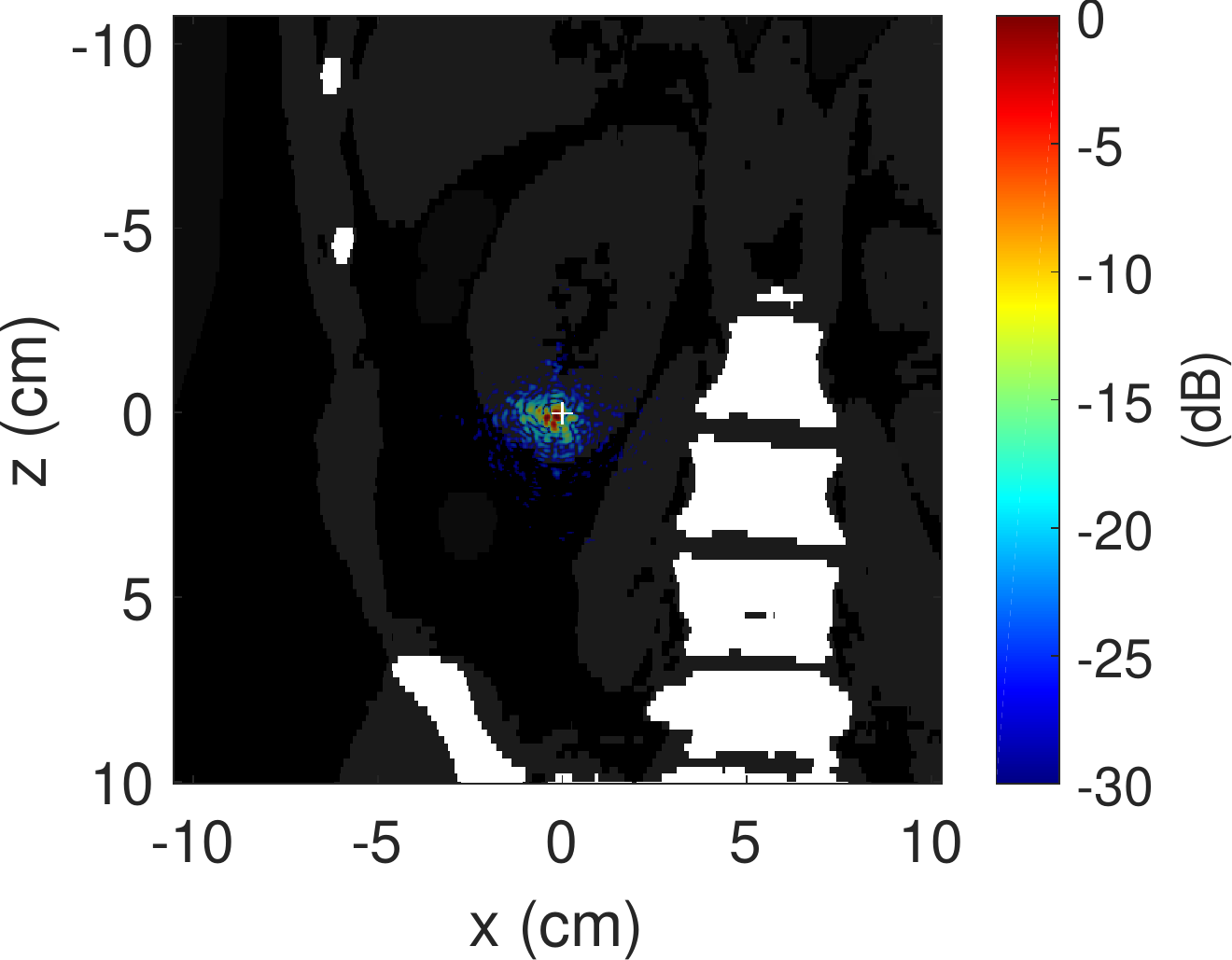}
    }
    \caption{(a) Axial, (b) sagittal and (c) coronal slices of the CT scan showing the ultrasound pressure field in kidney. The pressure field is displayed on a log-scale with a dynamic range of 30 dB. The different gray levels in the CT data correspond to the density of each tissue type: white - bone, gray - kidney/soft tissue, black - water/fat. The ultrasound focal point target location is marked with a white cross.}
    \label{fig:ct_us}
\end{figure*}

The simulation geometry was derived from a CT dataset of a patient (see Figure \ref{fig:ct_us}). Thresholds were used to segment the data set into bone, fat and other soft tissue. The kidney was then segmented manually. The medium outside the patient was assumed to be water. Typical values for sound speed, attenuation, density and B/A were used for each tissue type (see Table \ref{tab:tissue_properties}) \citep{mast2000empirical}. The HIFU transducer was modelled on a clinical system (Model JC-200 Tumor System, HAIFU) \citep{ritchie2013attenuation} with an annular transmitting surface of outer diameter 20 cm and inner hole diameter 6 cm. The operating frequency was 0.95 MHz and the focal length was 14.5 cm. The transducer was positioned so that the geometric focal point of the transducer (the white cross in Figure \ref{fig:ct_us}) was located in the bottom part of left kidney.

\begin{table}[b!]
  \centering
  \caption{Tissue parameters used in the simulations \citep{mast2000empirical}}
    \begin{tabular}{lcccc}
    \hline
          & Density   		& Sound speed   & Attenuation 							& B/A 	\\
          & (kg/m$^{3}$) 	& (m/s) 		& (dB/(MHz$^{\textrm{1.1}}\cdot$cm))	& 		\\
    \hline
    Water & 1000  			& 1520  		& 0.00217 							  	& 5.2 	\\
    Bone  & 1908  			& 4080  		& 20.00    							   	& 7.4 	\\
    Soft tissue & 1055  	& 1575  		& 0.60   							   	& 7.0 	\\
    Fat   & 950   			& 1478  		& 0.48  							   	& 10.0 	\\
    Kidney& 1050  			& 1560  		& 1.00     							   	& 7.4 	\\
    \hline
    \end{tabular}
  \label{tab:tissue_properties}
\end{table}

For data analysis three simulations were conducted: (i) reference simulation in pure water, (ii) simulation without the refraction effects (i.e., constant sound speed of water in all tissue types) but all other properties varying and (iii) simulation with all properties varying (i.e., with refraction effects). Before performing the actual simulations several convergence studies were conducted in order to find out the optimal grid size and temporal resolution. The computational grid consisted of 1200 $\times$ 1200 $\times$ 1200 grid points (i.e., 22.2 cm $\times$ 22.2 cm $\times$ 22.2 cm) giving a spatial resolution of 185 $\mu$m which supported nonlinear harmonic frequencies up to 4 MHz. Perfectly matched layers (PML) were used on the edges of the grid. The simulation length was set to 260 $\mu$s with a temporal resolution of 8.15 ns giving a total of 31876 time steps per simulation. The simulations were run using 400 computing cores for approximately 180 hours in total using the computing facilities provided by advanced research computing (ARC) at the University of Oxford \citep{richards2016arc}. For data analysis the time-domain waveforms and the peak pressures were saved in a three-dimensional grid around the focal point. In addition axial, sagittal and coronal slices of the ultrasound field over the whole spatial domain were saved.

\section*{Results}

Figure \ref{fig:ct_us} shows the axial, sagittal and coronal slices of the ultrasound pressure field generated by the HIFU transducer. The pressure field is displayed in log-scale  with a dynamic range of 30 dB.  The transducer was positioned in order to avoid the ribs which would otherwise cause significant pressure losses due to strong reflection. The annular nature of the source results in the appearance of two beams. In the focal region it can be seen that the region of high pressure does not form the archetypical ellipse shape, but is more diffuse. Further, the highest pressure is offset from the target location (white cross marker) in all slices. 

\begin{figure*}[htbp!]
    \vspace{2mm}
    \centering
    \subfigure[]
    {
        \includegraphics[width=0.3\textwidth]{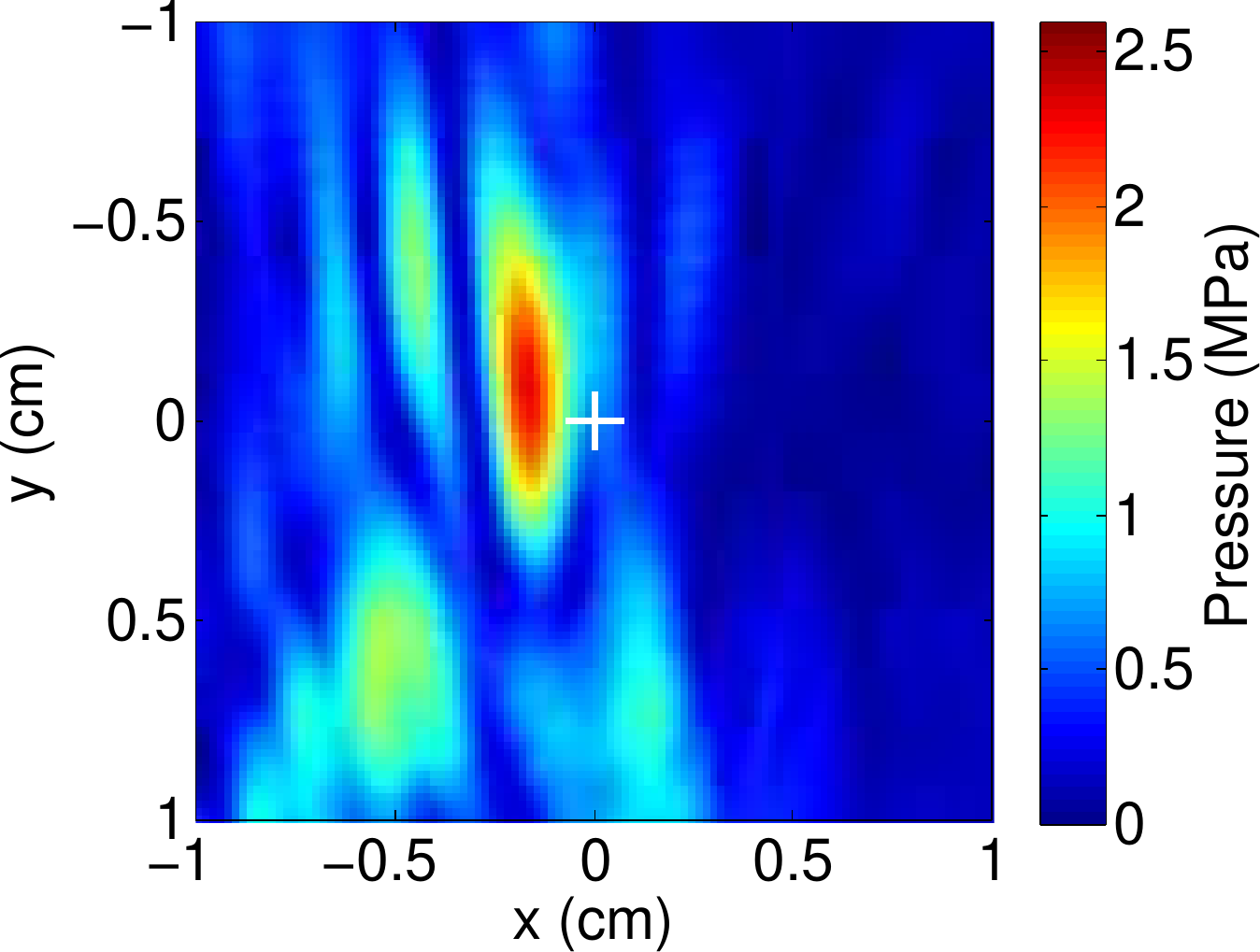}
    }
    \subfigure[]
    {
        \includegraphics[width=0.3\textwidth]{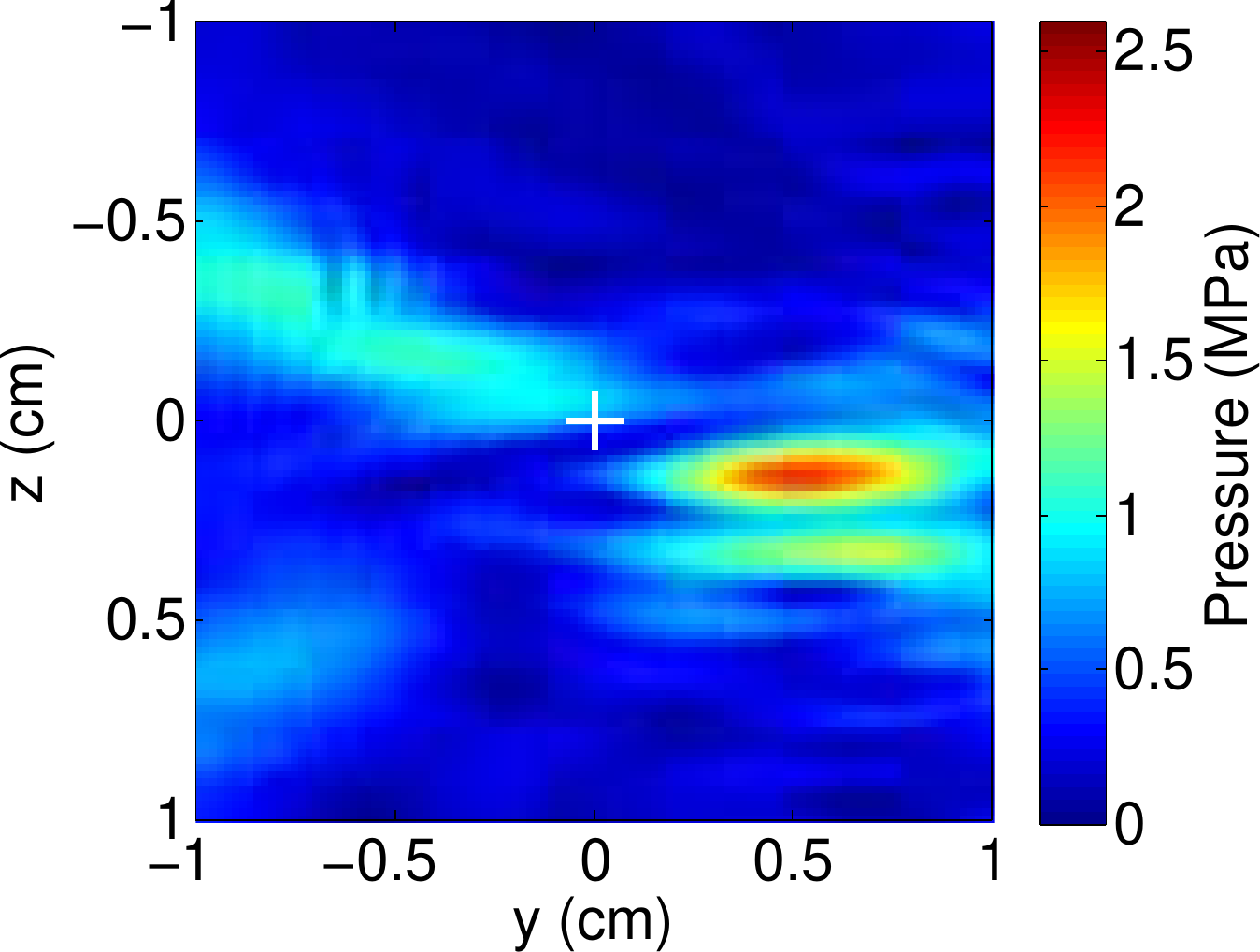}
    }
    \subfigure[]
    {
        \includegraphics[width=0.3\textwidth]{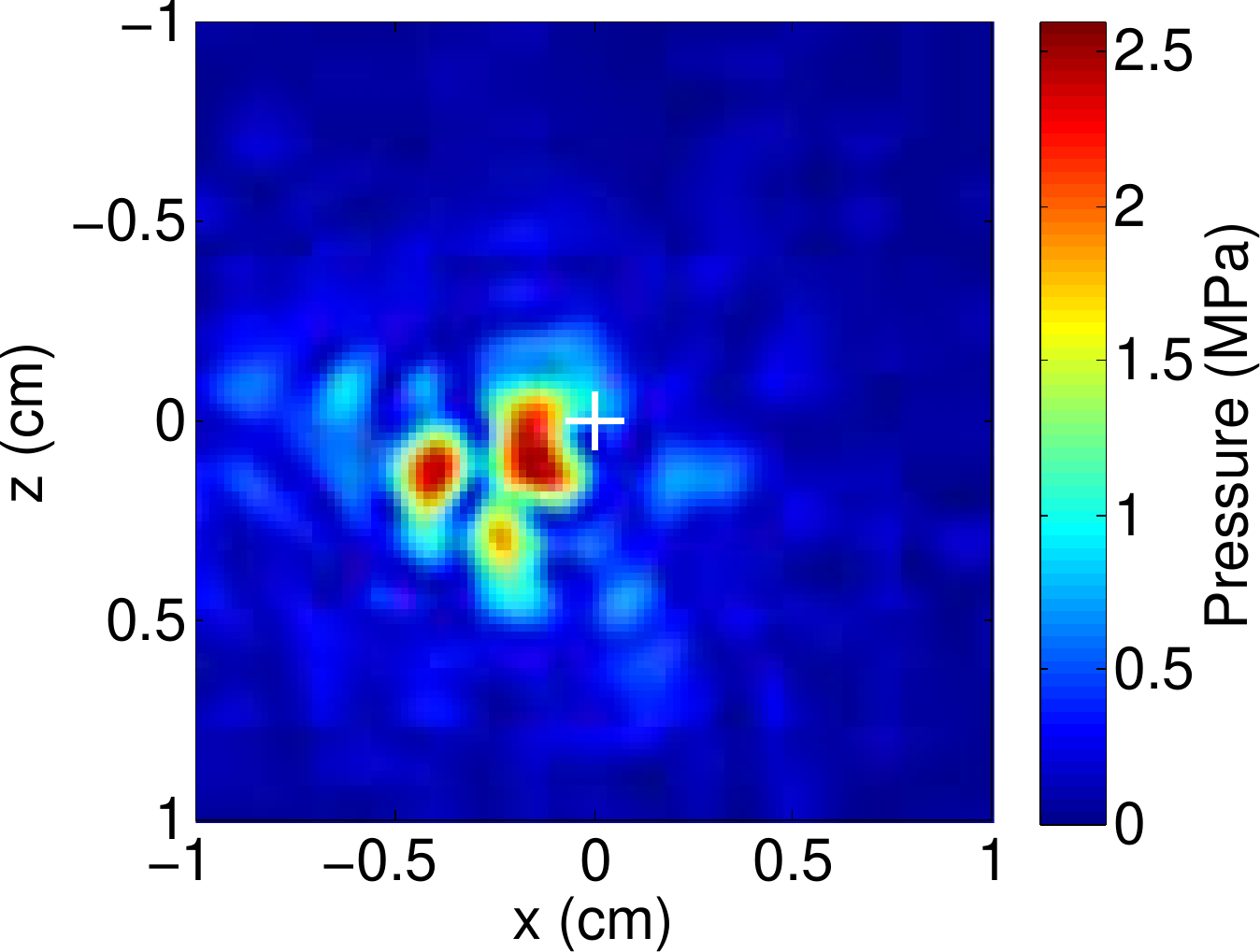}
    }
    \caption{(a) Axial, (b) sagittal and (c) coronal slices of the ultrasound field in the focal area in kidney. The ultrasound focal point target location is marked with a cross.}
    \label{fig:us_focal}
\end{figure*}

\begin{figure}[htbp!]
    \centering
    \subfigure[]
    {
        \includegraphics[width=0.43\columnwidth]{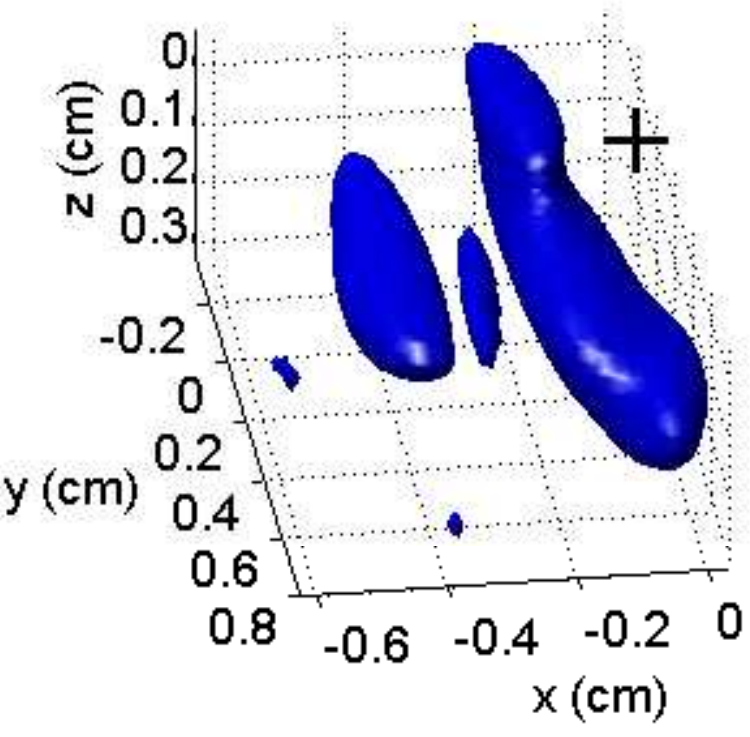}
    }
    \subfigure[]
    {
        \includegraphics[width=0.50\columnwidth]{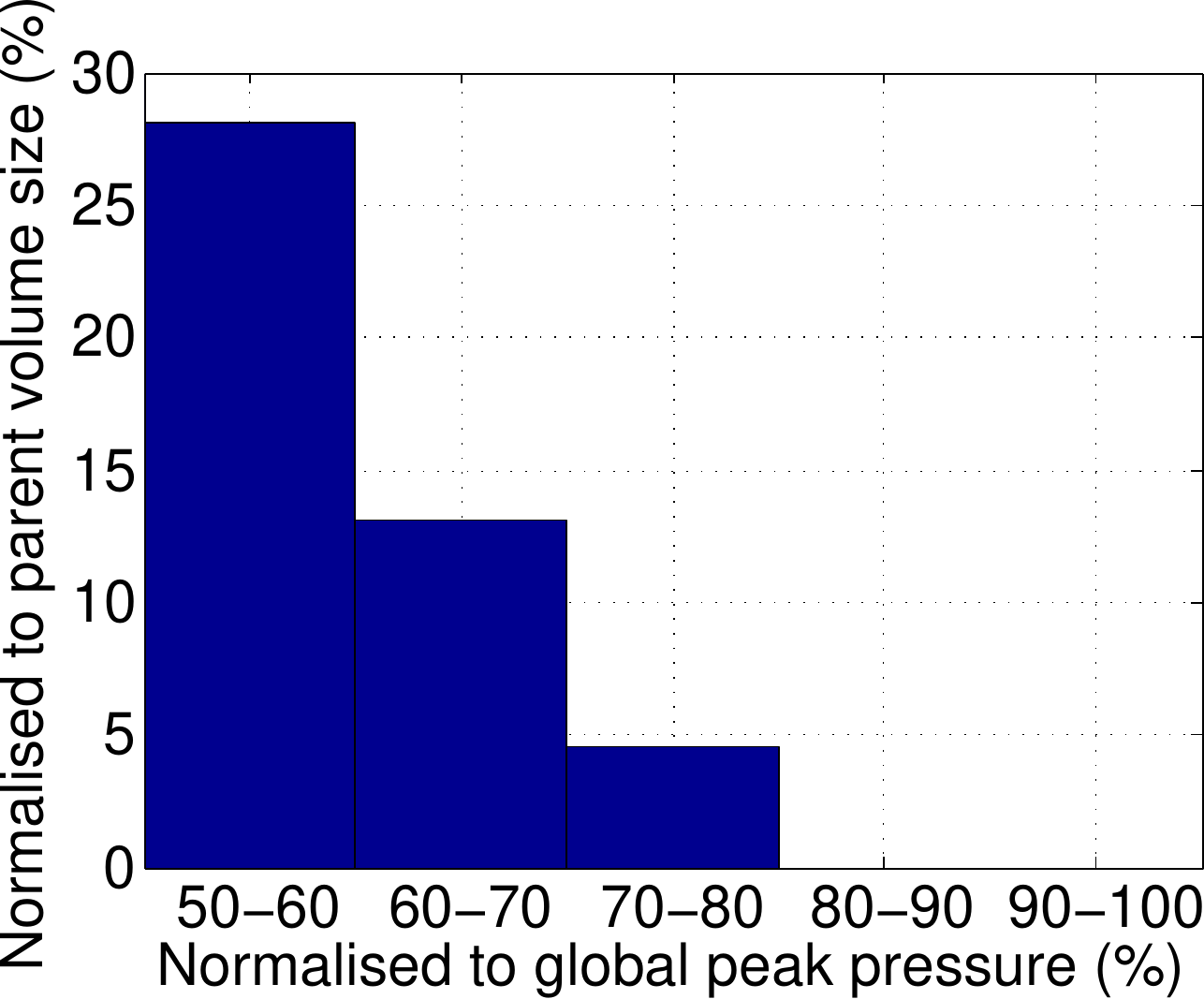}
    }
    \caption{(a) The focal point volume is shown with isosurfaces thresholded at -6 dB. The target focal point is marked with a black cross. The shifting and splitting of the focal point into one parent and four child focal volumes can be seen. (b) Histogram showing the size of the child volumes relative to the parent focal volume for various pressure contours varying from 50\% to 80\% of the global peak pressure.}
    \label{fig:focal_3D_hist}
\end{figure}

Figure \ref{fig:us_focal} shows close-ups of the axial, sagittal and coronal slices of the pressure field in the ultrasound focal area. Here the shift of the location of the highest pressure from the target location is clear and it was determined to be $-$1.3, 2.6 and 1.3 mm in x-, y- and z-directions respectively. By examining the focal area in more detail in the coronal slice (see Figure \ref{fig:us_focal}(c)), it can be seen that in addition to the focal shifting, a region of high pressure has split into a number of subvolumes. This is more clearly visualised in Figure \ref{fig:focal_3D_hist}(a) which shows the isosurfaces of the focal pressure regions thresholded at $-$6 dB. It can be seen that the focal region consists of five smaller focal points with the largest being approximately 12 mm in length and 3 mm in width. In comparison the size of the $-$6 dB focal point in water is approximately 6.5 mm in length and 1.1 mm in width.

The splitting of the focal region was quantified by identifying the largest subvolume as the parent focal region and the others as child regions. For a given pressure threshold, between 50\% and 100\% of the maximum pressure, the volume of the child focal regions was compared to that of the parent focal region. Figure \ref{fig:focal_3D_hist}(b) shows a histogram of the analysis. For pressure thresholds above 80\% no voxels were present in the child focal regions. However, when the threshold was reduced to 70\% it was found that approximately 5\% of the voxels were in the child focal regions. As the threshold was decreased the amount of volume in the child regions increased. At the $-$6 dB pressure threshold the total volume in the child regions was 46\% of the volume of the parent focal region. These data suggest that undesired heating effects will occur at secondary focal points due to focal point splitting.

\begin{figure}[htbp!]
    \centering
    \subfigure[]
    {
        \includegraphics[width=0.46\columnwidth]{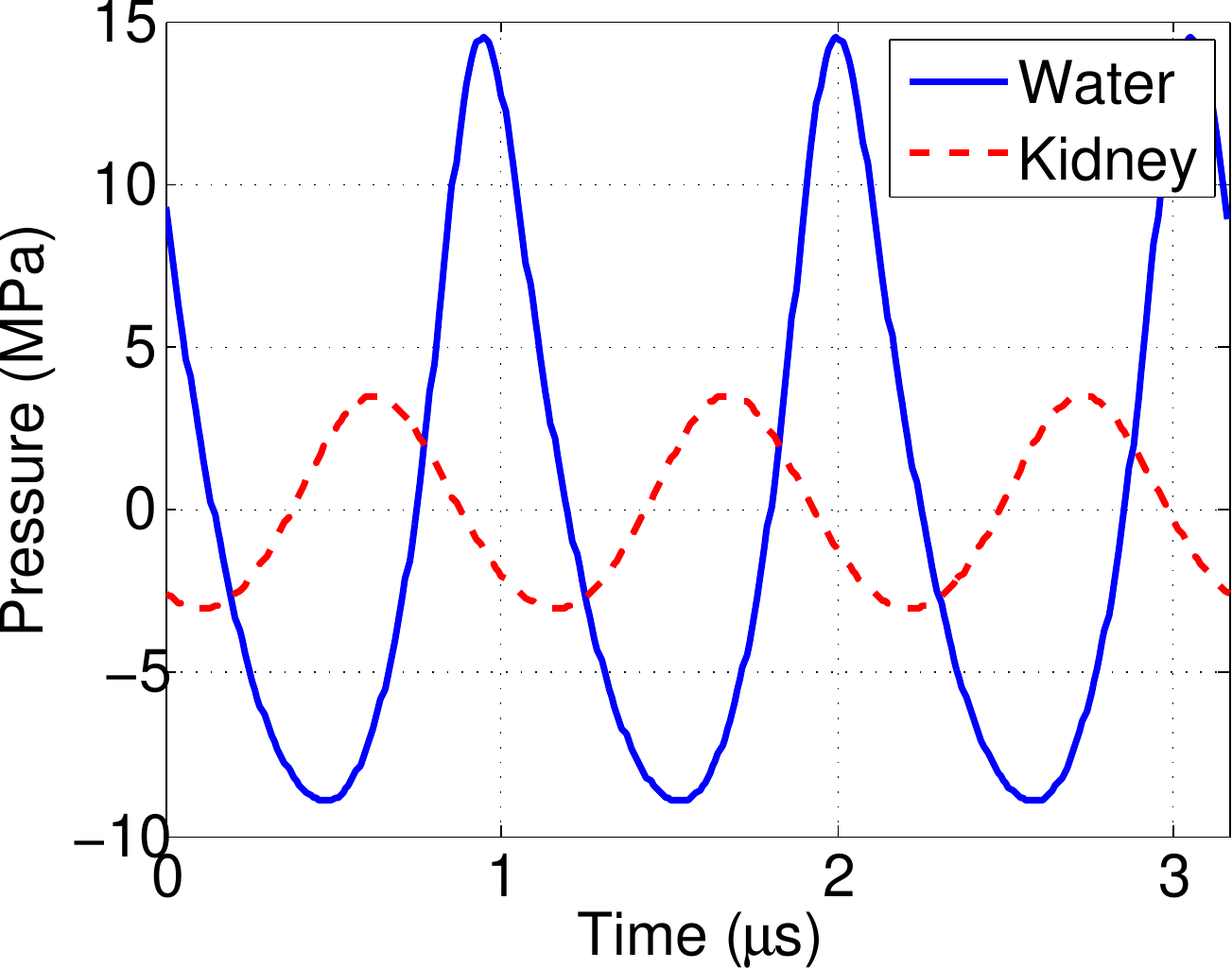}
    }
    \subfigure[]
    {
        \includegraphics[width=0.46\columnwidth]{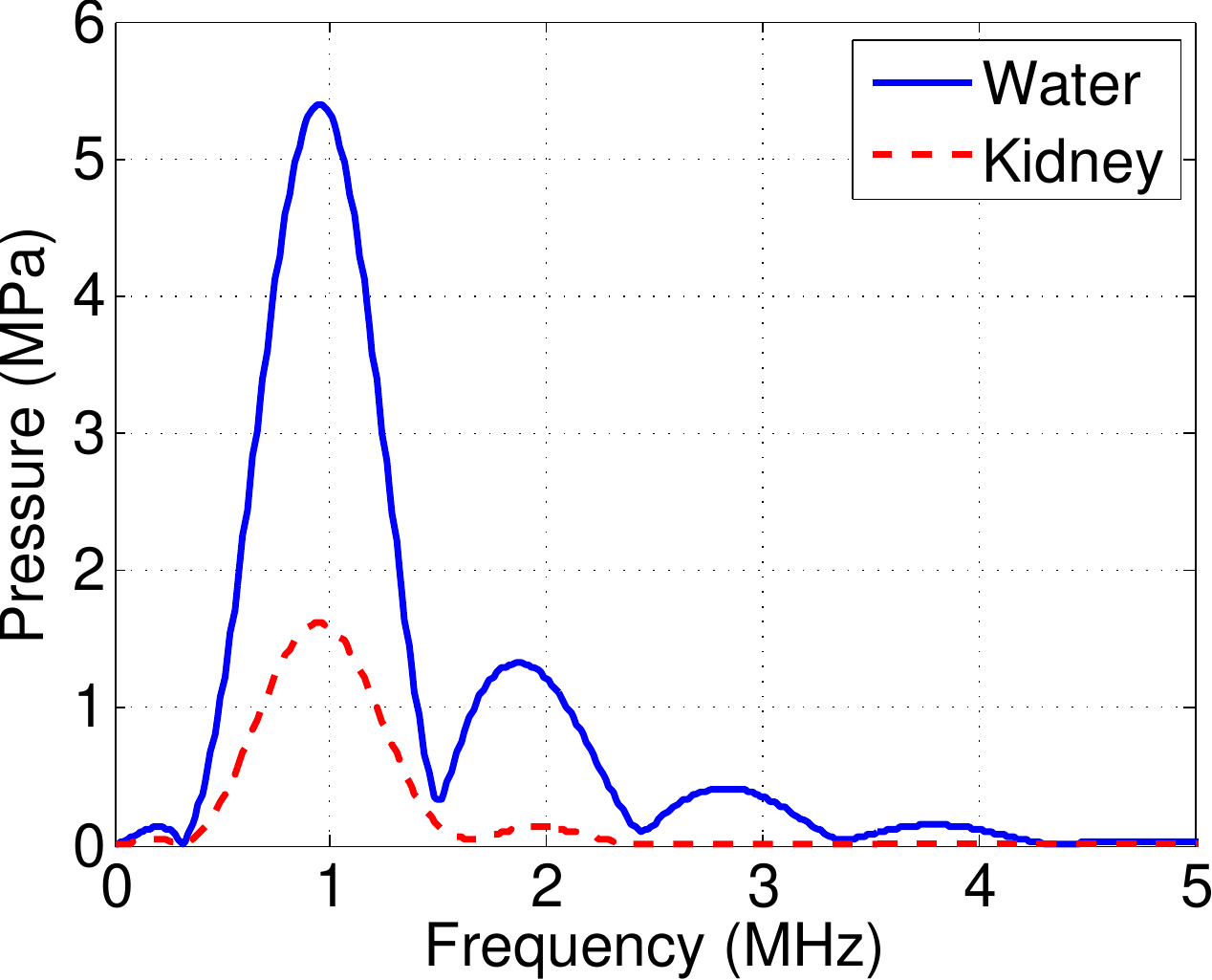}
    }
    \caption{(a) Time domain waveforms at the maximum peak pressure location in water and kidney. (b) Windowed (Hann) frequency spectrum of the same waveforms.}
    \label{fig:waveform_time_fft}
\end{figure}

Figure \ref{fig:waveform_time_fft}(a) shows the time waveforms at the location of maximum peak pressure in both water and kidney. The peak-positive pressure drops from 14.49 MPa in water to 3.51 MPa in kidney. Similarly, the spatial peak-temporal average (SPTA) intensity has dropped from 4116 W/cm$^{2}$ in water to 324 W/cm$^{2}$ in kidney, that is, an 11.0 dB decrease. The simulation without the refraction effects resulted in a single focal point (i.e., no focal splitting) with a peak-positive pressure of 6.46 MPa and SPTA intensity of 957 W/cm$^{2}$ corresponding to a 6.3 dB decrease. This suggests that refraction and attenuation contribute similarly to the loss in focal intensity. Figure \ref{fig:waveform_time_fft}(b) shows the windowed (Hann) frequency spectrum of the same signals. A peak at centre frequency 0.95 MHz is clearly visible as are the nonlinearly generated harmonics, however, in the case of tissue the nonlinear effects are much less pronounced.

\section*{Discussion}

\citet{ritchie2013attenuation} studied the attenuation of focused ultrasound using subcutaneous and peri-nephric fat layers in front of the HIFU transducer. They found the attenuation of peri-nephric fat to be significantly higher (1.36 dB/cm) compared to typical fat tissue attenuation (0.48 dB/cm) \citep{mast2000empirical}. In the simulations reported here all the fat layers were segmented as normal fat tissue using the latter attenuation value. This difference is not thought to affect conclusions as for the patient derived data set employed here the thickness of peri-nephric fat was 0.5 cm on average and adding in the higher attenuation would contribute an extra 0.44 dB of loss which is small in comparison to the total loss observed. The most significant attenuation losses were caused by subcutaneous fat and soft tissue in front of the kidney whose thickness were approximately 2.6 and 3.7 cm respectively.

In addition to attenuation, energy losses also occur due to reflections and scattering at interfaces, such as, the rib cage, tissue interfaces and air pockets.  Here the transducer was positioned so that reflections due to rib bones were not present.  The effect of tissue interfaces in the penetration of HIFU has been studied in rabbit kidney \textit{in vivo} by \citet{damianou2004mri}. They found the ultrasound penetration through muscle-kidney and fat-kidney interfaces to be excellent in a situation where no air bubbles were present. However, in some cases air spaces existed in between these interfaces which caused strong reflections and acted as possible sites for cavitation during the HIFU therapy. In the simulations here the interfaces between tissues contained no air spaces, and therefore, all the possible energy losses due to reflections were caused either by the rib cage or acoustic impedance mismatches between tissue interfaces. The intensity transmission coefficients for water-fat, fat-soft tissue, soft tissue-fat and fat-kidney interfaces were 99.84\%, 99.29\%, 99.29\% and 99.41\% respectively.  For all the interfaces the total transmission is 97.85\% which corresponds to a loss of less than 0.1 dB.

Focal shifting and splitting due to variations in the sound speed is another factor considerably affecting the efficacy of HIFU therapy. At interfaces changes in sound speed result in refraction, in addition, the phase accumulation will change in different tissues affecting the constructive and destructive interference of the waves. These effects will impact both the intensity and the location of the focus. Focal shifting due to subcutaneous and peri-nephric fat was studied by \citet{ritchie2013attenuation} who found the shift to be approximately 1 mm in both transverse directions. In the simulations here similar magnitude shifts were observed which are large in terms of $-$6 dB focal point width (1.1 mm), but not with respect to typical renal tumour sizes of several centimetres \citep{remzi2007renal}. Splitting of the ultrasound focal point into smaller, less defined, volumes can significantly reduce its heating efficiency. The simulations showed an 11.0 dB drop in SPTA intensity when all effects were incorporated (specifically attenuation and refraction) and only a 6.3 dB drop without the refraction effects (i.e., no focal point splitting). This suggests that attenuation and refraction have a similar impact on the intensity loss at the focus, contributing about 5 to 6 dB each. When focal point splitting was present, the cumulative size of the two separate smaller focal volumes was found to be approximately 46\% of main focal point. Although focal splitting provides larger total heating volume the efficiency is reduced, because the acoustic energy is distributed over a larger volume which leads to longer therapy times and also may result in undesired heating in regions away from the target region.

Other phenomena that have been shown to reduce the efficacy of renal HIFU therapy are respiratory movement \citep{marberger2005extracorporeal} and perfusion \citep{chang2004effects}.  These effects were not incorporated in the simulation model but could be considered in the future.

\section*{Conclusions}

The effects of attenuation, reflection and refraction on the efficacy of HIFU therapy in kidney were investigated using a nonlinear simulation model. Attenuation and splitting due to refraction were found to be the most significant factors reducing the intensity of the ultrasound field. Reflections due to the rib cage could possibly cause significant losses, but this can be avoided by optimal positioning of the transducer. The reflections due to tissue interfaces were found to be negligible.
        
\section*{Acknowledgements}

V.~S. acknowledges the support of the RCUK Digital Economy Programme grant number EP/G036861/1 (Oxford Centre for Doctoral Training in Healthcare Innovation) as well as the support of Instrumentarium Science Foundation, Jenny and Antti Wihuri Foundation and Finnish Cultural Foundation. J.~J. is financed from the SoMoPro II Programme, co-financed by the European Union and the South-Moravian Region. This work reflects only the author's view and the European Union is not liable for any use that may be made of the information contained therein. B.~T. acknowledges the support of EPSRC grant numbers EP/L020262/1 and EP/M011119/1. R.~C. acknowledges the support of EPSRC grant number EP/K02020X/1.





\pagebreak

\bibliographystyle{UMB-elsarticle-harvbib}
\bibliography{arXiv_embc}

\end{document}